\documentclass[prx,twocolumn,floatfix,superscriptaddress,aps,longbibliography]{revtex4-1} 
\pdfoutput=1

\usepackage{amsfonts,amsmath,amssymb,color,times,graphicx}
\definecolor{darkblue}{rgb}{0, 0, 0.8}
\usepackage[colorlinks=true, breaklinks=true, linkcolor=darkblue, citecolor=darkblue, urlcolor=darkblue]{hyperref}

\usepackage{graphicx}
\usepackage{bmpsize}
\usepackage{amsfonts}
\usepackage{gensymb}
\usepackage{braket}
\usepackage{mathtools}
\usepackage{bm}

\begin{document}
\title{Neutron Scattering Studies on the High-$T_c$ Superconductor La$_3$Ni$_2$O$_{7-\delta}$ at Ambient Pressure}

\author{Tao Xie}
\author{Mengwu Huo}
\author{Xiaosheng Ni}
\affiliation{Center for Neutron Science and Technology, Guangdong Provincial Key Laboratory of Magnetoelectric Physics and Devices, School of Physics, Sun Yat-Sen University, Guangzhou, Guangdong 510275, China}
\author{Feiran Shen}
\affiliation{Spallation Neutron Source Science Center, Dongguan, Guangdong 523803, China}
\author{Xing Huang}
\affiliation{Center for Neutron Science and Technology, Guangdong Provincial Key Laboratory of Magnetoelectric Physics and Devices, School of Physics, Sun Yat-Sen University, Guangzhou, Guangdong 510275, China}

\author{Hualei Sun}
\affiliation{School of Science, Sun Yat-Sen University, Shenzhen, Guangdong 518107, China}

\author{Helen C. Walker}
\affiliation{ISIS Neutron and Muon Source, Science and Technology Facilities Council, Rutherford Appleton Laboratory, Didcot OX11 0QX, United Kingdom}

\author{Devashibhai Adroja}
\affiliation{ISIS Neutron and Muon Source, Science and Technology Facilities Council, Rutherford Appleton Laboratory, Didcot OX11 0QX, United Kingdom}

\author{Dehong Yu}
\affiliation{Australian Nuclear Science and Technology Organisation, Locked bag 2001, Kirrawee DC, NSW 2232, Australia}

\author{Bing Shen}
\affiliation{Center for Neutron Science and Technology, Guangdong Provincial Key Laboratory of Magnetoelectric Physics and Devices, School of Physics, Sun Yat-Sen University, Guangzhou, Guangdong 510275, China}

\author{Lunhua He}
\affiliation{Beijing National Laboratory for Condensed Matter Physics, Institute of Physics, Chinese Academy of Sciences, Beijing, 100190, China}
\affiliation{Spallation Neutron Source Science Center, Dongguan, Guangdong 523803, China}
\affiliation{Songshan Lake Materials Laboratory, Dongguan, Guangdong 523808, China}

\author{Kun Cao}
\author{Meng Wang}
\thanks{Corresponding author: wangmeng5@mail.sysu.edu.cn}
\affiliation{Center for Neutron Science and Technology, Guangdong Provincial Key Laboratory of Magnetoelectric Physics and Devices, School of Physics, Sun Yat-Sen University, Guangzhou, Guangdong 510275, China}


\begin{abstract}
After several decades of studies of high-temperature superconductivity, there is no compelling theory for the mechanism yet; however, the spin fluctuations have been widely believed to play a crucial role in forming the superconducting Cooper pairs. The recent discovery of high-temperature superconductivity near 80 K in the bilayer nickelate La$_3$Ni$_2$O$_7$ under pressure provides a new platform to elucidate the origins of high-temperature superconductivity. We perform elastic and inelastic neutron scattering studies on a polycrystalline sample of La$_3$Ni$_2$O$_{7-\delta}$ at ambient pressure. No magnetic order can be identified down to 10 K. The absence of long-range magnetic order in neutron diffraction measurements may be ascribed to the smallness of the magnetic moment. However, we observe a weak flat spin-fluctuation signal at $\sim$ 45 meV in the inelastic scattering spectra. The observed spin excitations could be interpreted as a result of strong interlayer and weak intralayer magnetic couplings for stripe-type antiferromagnetic orders. Our results provide crucial information on the spin dynamics and are thus important for understanding the superconductivity in La$_3$Ni$_2$O$_7$.

\noindent{\bf Keywords: high-$T_c$ superconductor, bilayer nickelate La$_3$Ni$_2$O$_7$, neutron scattering, spin excitations.}
\end{abstract}

\maketitle

\textit{Introduction}.
The mechanism of high-temperature (high-$T_c$) superconductivity is a long-lasting mystery although tremendous progress has been achieved since the discovery of superconductivity in cuprates and iron-based compounds. High-$T_c$ superconductivity in copper-oxide and iron-based superconductors always appears after the suppression of the antiferromagnetic (AFM) order in the parent compounds by chemical doping or pressure and evolves into a superconducting dome with the highest $T_c$ around the critical doping/pressure where the AFM order disappears~\cite{Scalapino_RevModPhys,Tranquada2014,Keimer_Nature,Dai_RevModPhys}. As the AFM order is suppressed by chemical doping, the spin fluctuations persist throughout the phase diagram and form a neutron spin resonant mode below $T_c$ at certain energies around the AFM wave vectors of the parent compounds. The temperature dependence of the intensity of the neutron spin resonant mode behaves like a superconducting order parameter and the resonant energy can be linearly scaled with the corresponding $T_c$ and superconducting gap~\cite{Eschrig2006,Sidis2007,Christianson2008resonance,Yu_NPhys}. All these characteristics suggest that magnetism is crucial to the mechanism of the high-$T_c$ superconductivity, and the AFM spin fluctuations could be the ``glue" for the formation of superconducting Cooper pairs.

Seeking new unconventional high-$T_c$ superconductors is the key to understanding the high-$T_c$ mechanism and extending the application of superconductivity. Nickelates have been proposed to be a promising material platform to search for high-$T_c$ superconductors due to the similarity of the lattice and electronic structure with those of the cuprates~\cite{Anisimov1999}. Among different families of nickelates, $RE$NiO$_2$ ($RE$ = La, Nd, Pr) were predicted to be the most probable to achieve superconductivity due to the same electronic configuration of Ni$^{+}$ (3$d^9$) with Cu$^{2+}$ in the cuprates. Superconductivity was indeed observed in Nd$_{0.8}$Sr$_{0.2}$NiO$_2$ thin films ($T_c$ $\sim$ 9-15 K) in 2019 and later confirmed in other thin film samples with similar compositions~\cite{Lidanfeng_Nature,Osada2020,Wang_NatCom,GU112review}. Additionally, it is interesting to note that spin-wave-like magnetic fluctuations of Nd$_{1-x}$Sr$_{x}$NiO$_2$ from resonant inelastic X-ray scattering (RIXS) measurements give comparable magnetic exchange couplings to that in copper-oxide superconductors~\cite{Lu_NdNiO2,Coldea_La2CuO4}.

\begin{figure}[t]
\center{\includegraphics[width=1\linewidth]{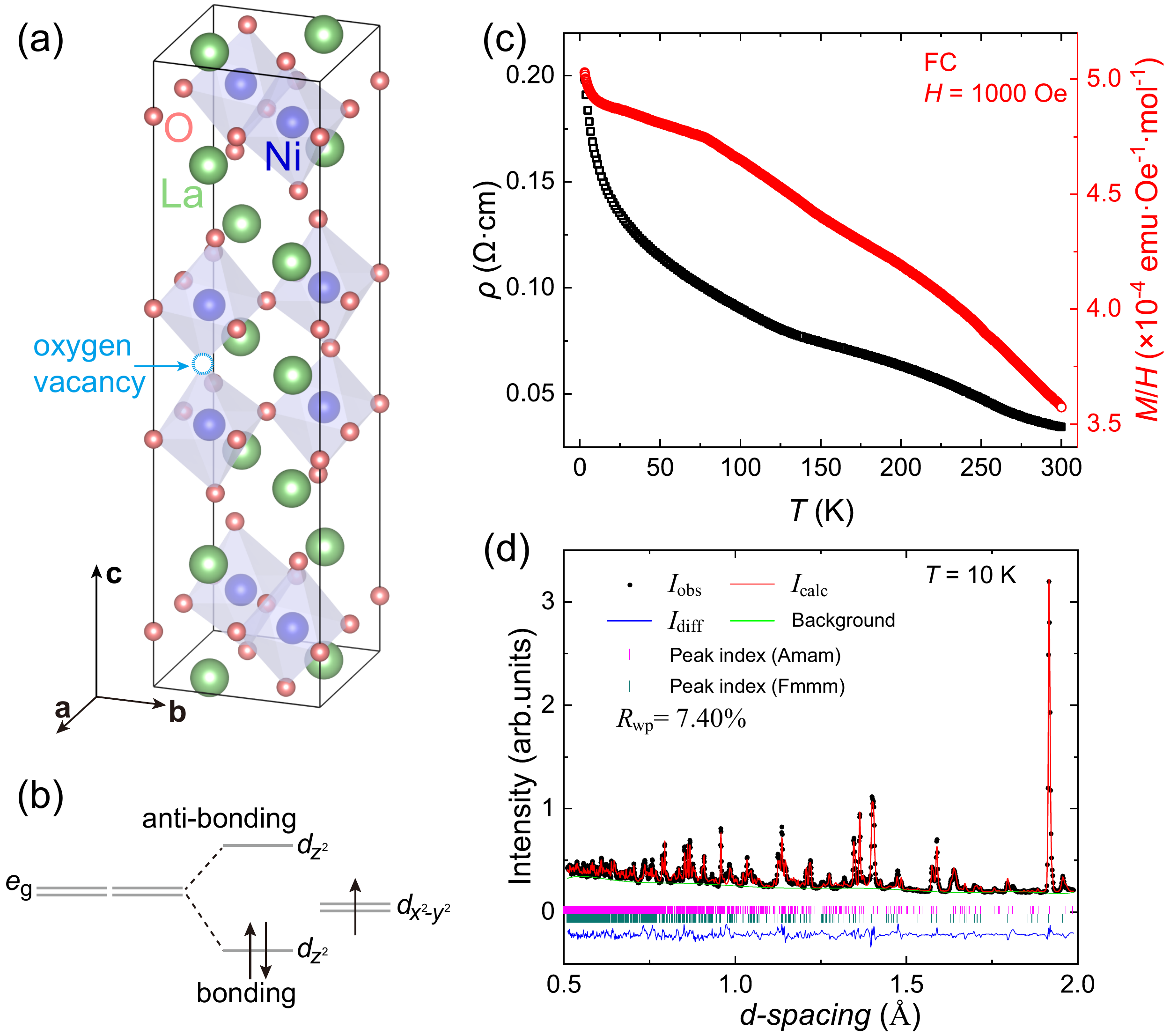}}
\caption{{Structure, electronic occupancy, and bulk electrical transport, magnetic and structural characterization of the La$_3$Ni$_2$O$_{7-\delta}$ polycrystalline sample.} (a) Crystal structure of La$_3$Ni$_2$O$_{7-\delta}$ in the $Amam$ space group. The open blue circle represents an oxygen vacancy.
(b) Schematic of the $\sigma$-bonding and anti-bonding states for two Ni$^{2.5+}$ (3$d^{7.5}$).
(c) Temperature dependence of electrical resistivity and field cooled (FC) magnetization ($H$ = 1000 Oe).
(d) Neutron powder diffraction pattern collected at 10 K. }
\label{fig1}
\end{figure}

Recently, superconductivity with $T_c$ up to 80 K was reported in single-crystal samples of the Ruddlesden-Popper phase bilayer nickelate La$_3$Ni$_2$O$_7$ [Fig.~\ref{fig1}(a)] with pressure above 14 GPa~\cite{SunHL_Nature}. Either zero-resistance or a diamagnetic response corresponding to the superconducting transition in La$_3$Ni$_2$O$_7$ has been confirmed on single-crystal or polycrystalline samples in subsequent reports~\cite{Hou2023,Zhang2023High,Wang2023pressure,Zhou}. X-ray photoemission spectroscopy measurements indicated the valence state of the Ni ions in La$_3$Ni$_2$O$_7$ is a mixture of Ni$^{2+}$ and Ni$^{3+}$, which may result in a charge density wave (CDW) and a spin density wave (SDW)\cite{Liu2023,Taniguchi1995}. Electrical transport and specific heat measurements on single crystal samples indeed revealed transition-like anomalies in resistance and heat capacity at 153 and 110 K~\cite{Liu2023}. An optical study observed a CDW-like transition at $\sim$115 K~\cite{Liu2023wen}. Recent $\mu$SR, RIXS and nuclear magnetic resonance (NMR) studies provided evidences of the existence of the SDW order in a single-crystal and powder La$_3$Ni$_2$O$_{7-\delta}$ below $\sim$150 K~\cite{Chen2023musR,Khasanov2024musR,Chen2024RIXS,Dan2024NMR}. However, neutron diffraction measurements did not observe long-range magnetic order in powder samples~\cite{Ling1999}. Thus, an elucidation of the nature of the magnetic ground state, the properties of the spin fluctuations, and the magnitudes of the magnetic exchange interactions of La$_3$Ni$_2$O$_7$ are crucial to understanding the mechanism of the high-$T_c$ superconductivity.

Although the angle of the bilayer Ni-O-Ni along the $c$ axis evolves from 168$^{\circ}$ to 180$^\circ$ under pressure~\cite{SunHL_Nature}, the spin fluctuations and magnetic exchange interactions under ambient conditions should be comparable to those at high pressure.
In this paper, we report neutron diffraction and systematic inelastic neutron scattering (INS) measurements on a polycrystalline La$_3$Ni$_2$O$_{7-\delta}$ sample at ambient pressure. No long-range magnetic order can be identified at 10 K. The INS data reveal a non-dispersive spin excitation signal at $\sim45$~meV, which has been used to estimate the magnitude of the magnetic exchange couplings with stripe-type (the single spin-charge stripe or the double spin stripe) antiferromagnetic orders. In addition, dispersive and non-dispersive phonon modes are observed. By employing density functional theory (DFT) calculations, we have computed the dispersion and the density of states (DOS) of the phonon modes, which are roughly consistent with the INS spectra. Our results reveal that the strong interlayer magnetic exchange couplings dominate the spin fluctuations, which is distinct from the situation for the cuprates and iron-based superconductors.

\begin{figure}[t]
\center{\includegraphics[width=1\linewidth]{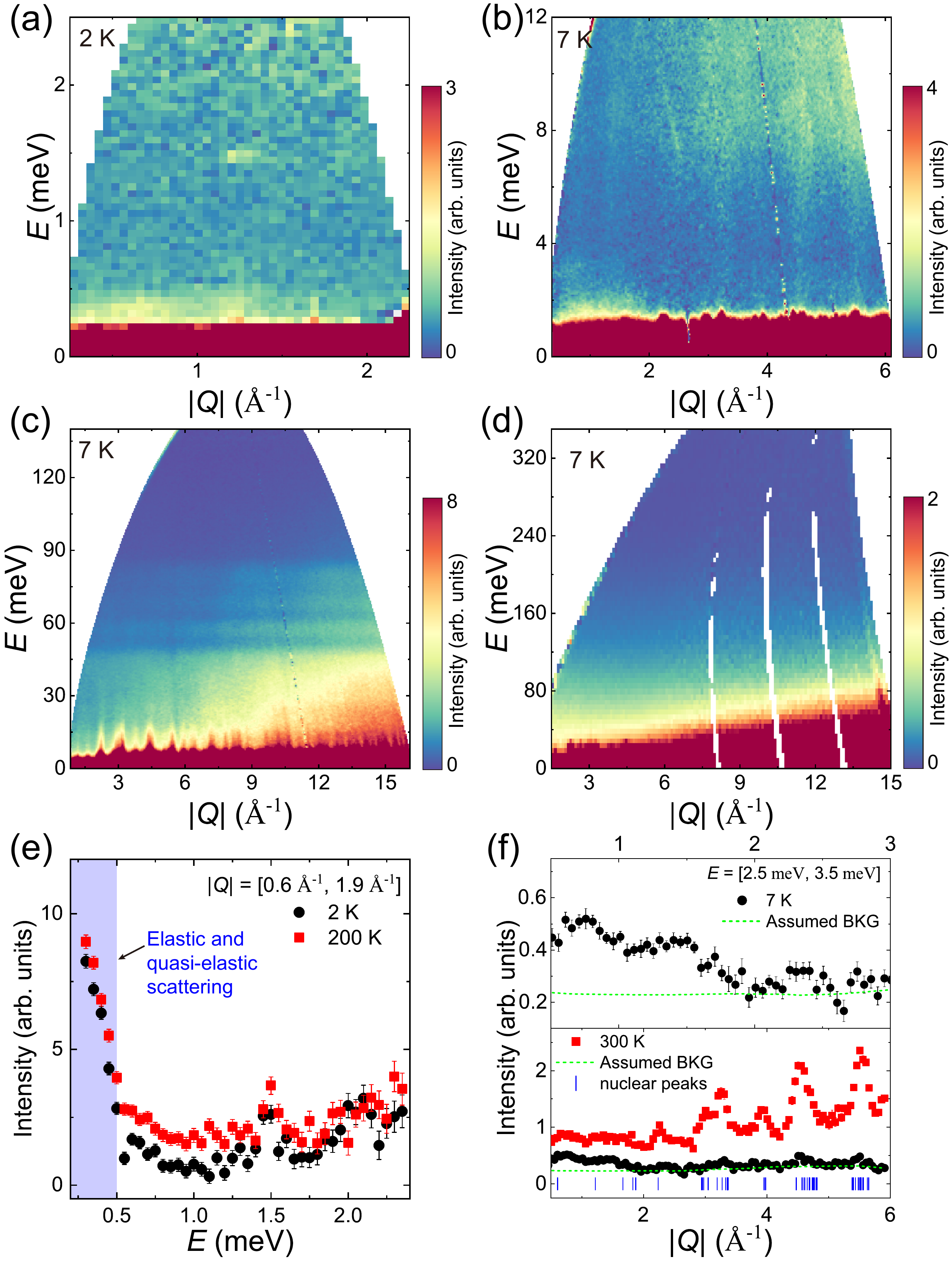}}
\caption{{INS spectra of La$_3$Ni$_2$O$_{7-\delta}$ with different incident energies at low temperature.}
(a) INS spectrum collected at PELICAN with $E_i$ = 3.7 meV. (b) and (c) are INS spectra measured at MERLIN with $E_i$ = 22.6 and 160 meV, respectively. (d) INS spectrum up to 350 meV obtained at MAPS. (e) A constant $|Q|$ cut in the range of [0.6, 1.9] {\AA}$^{-1}$ for $E_i$ = 3.7 meV. The shadow region indicates the elastic and quasi-elastic scatterings from the sample. The peak at 1.5 meV is from spurious scattering which can also be observed in panel (a). (f) Constant energy cuts within $E$ = [2.5, 3.5] meV at 7 K (the upper panel) and a comparison between 7 and 300 K (the lower panel) for $E_i$ = 22.6 meV.}
\label{fig2}
\end{figure}

\begin{figure*}[t]
\center{\includegraphics[width=0.9\linewidth]{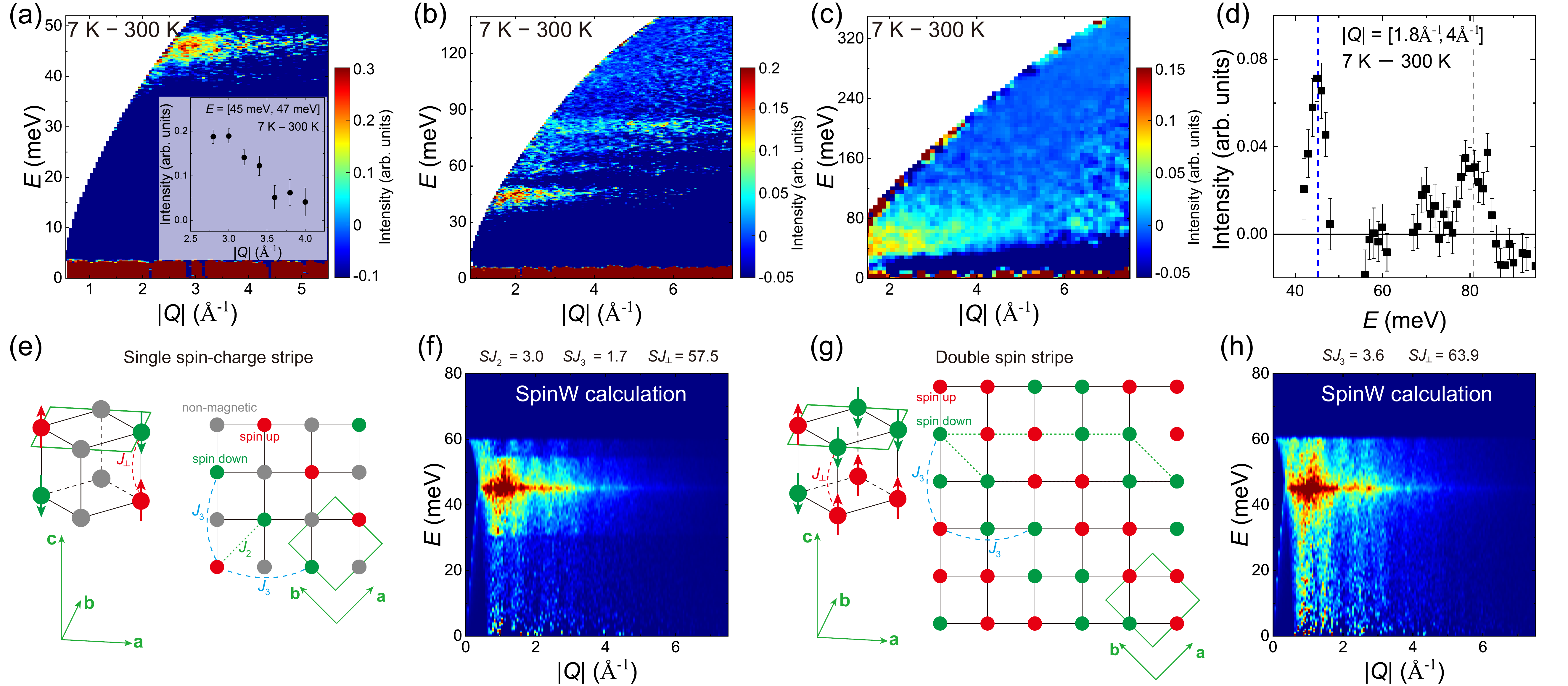}}
\caption{{INS spectra after subtraction of high from low temperature data.}
(a)-(c) show the resultant data between 7 and 300 K for $E_i$s = 66.3, 160, and 500 meV. The inset of panel (a) is a constant energy cut with $E$ = [45, 47] meV. (d) Constant $|Q|$ cut from (b) with $|Q|$ = [1.8, 4] ~{\AA}$^{-1}$. The dashed lines mark the centers of the peaks. (e) The single spin-charge stripe AFM order used for the SpinW calculations. The green and red balls represent the magnetic Ni atoms with spin up and down, respectively. The gray balls represent the nonmagnetic (charge-only) Ni atoms. The green solid lines indicate the unit cell of the orthorhombic $Amam$ structure. The in-plane exchange couplings $J_2$, $J_3$, and the inter-layer exchange coupling $J_{\perp}$ are indicated in the structure. (f) The calculated spin excitation spectrum from \textsc{SpinW} with $SJ_2$ = 3.0, $SJ_3$ = 1.7, and $SJ_{\perp}$ = 57.5 meV. (g) The double spin stripe AFM order used for the SpinW calculations. The green and red balls represent the magnetic Ni atoms with spin up and down, respectively. The green solid lines indicate the unit cell of the orthorhombic $Amam$ structure. The in-plane exchange couplings $J_3$ and the inter-layer exchange coupling $J_{\perp}$ are indicated in the structure. (h) The calculated spin excitation spectrum from \textsc{SpinW} with $SJ_3$ = 3.6, and $SJ_{\perp}$ = 63.9 meV.}
\label{fig3}
\end{figure*}

\textit{Characterizations and experimental details}.
Figure~\ref{fig1}(c) shows basic bulk characterization of the sample. The temperature dependence of the resistivity goes up slightly at low temperatures. The magnetization at $H$ = 1000 Oe also increases towards lower temperatures. Both upward trends at low temperatures suggest the existence of oxygen vacancies~\cite{Zhang1994}. Neutron diffraction experiments were performed on the general purpose powder diffractometer (GPPD)~\cite{GPPD} at China Spallation Neutron Source (CSNS) at 10 and 160 K. The data were refined with the Rietveld method using the GSAS software package~\cite{GSAS1}. The low-energy INS spectra at 2 and 200 K were collected on the time-of-flight cold neutron spectrometer PELICAN~\cite{PELICAN} at the Australian Centre for Neutron Scattering (ACNS) at the Australian Nuclear Science and Technology Organisation (ANSTO) with an incident neutron energy $E_i$ = 3.7 meV. The high-energy spectra were obtained from the time-of-flight spectrometers MERLIN~\cite{MERLIN} and MAPS~\cite{MAPS} at the ISIS Neutron and Muon Source at UK with $E_i$s = 15.5, 22.6, 36, 66.3, 160, and 500 meV (MAPS) at 7 and 300 K. We used 5 g powder samples of La$_3$Ni$_2$O$_7$ synthesized by the solid-solid reaction method~\cite{Liu2023} for the experiments on GPPD and PELICAN. 20 g powder samples synthesized in the same way were used for the experiments on MERLIN and MAPS. The INS data from PELICAN were analyzed using the LAMP software package. The data collected at MERLIN and MAPS were reduced and analyzed using the software packages MANTIDPLOT~\cite{Mantid} and DAVE~\cite{DAVE}. The scattering intensities from aluminum have been removed by subtracting the data from an empty can for the PELICAN and MERLIN experiments. The spin excitation simulations were performed using the \textsc{SpinW} software package~\cite{Spinw}. The phonon dispersions and DOS were calculated using the finite displacement method as implemented in the Phonopy~\cite{Togo2015first} software in combination with the Vienna $ab\ initio$ Simulation Package (VASP)~\cite{Kresse1993,Kresse1996}.

\textit{Neutron diffraction results}. Figure~\ref{fig1}(d) presents the neutron powder diffraction data measured at 10 K. No extra peaks can be seen when comparing the data between 10 and 160 K (see supplementary information (SI)~\cite{SI} for details). Our refinements reveal two structures of La$_3$Ni$_2$O$_7$ with space groups of $Amam$ (No. 63) and $Fmmm$ (No. 69). The refined compositions at 10 K for the two phases are 56.1\% La$_3$Ni$_2$O$_{6.79}$ and 41.0\% La$_3$Ni$_2$O$_{6.90}$ for the $Amam$ and $Fmmm$ structures, respectively~\cite{SI}. A fraction of 2.9\% La$_2$NiO$_4$ impurity is also identified. Based on the refined results, there are more vacancies on the inner apical oxygen sites for the $Amam$ phase, consistent with a recent transmission electron microscopy experiment~\cite{Dong2023}. We thus use La$_3$Ni$_2$O$_{7-\delta}$ to refer to our sample hereafter. It should be noted that La$_3$Ni$_2$O$_7$ with the $Fmmm$ structure has also been reported at ambient pressure~\cite{Zhang1994}.

\textit{Inelastic neutron scattering results}.
Figures~\ref{fig2}(a)-\ref{fig2}(d) present the INS spectra measured with different $E_i$s at 2 K on PELICAN and 7 K on MERLIN and MAPS. Identical measurements at 200 and 300 K are presented in the SI~\cite{SI}. One cannot see obvious spin fluctuations from the color maps at low $|Q|$. At higher $|Q|$ and energy ranges weak acoustic phonon modes in Fig. \ref{fig2}(b) and flat optical phonon modes in Fig. \ref{fig2}(c) can be observed.
To inspect the INS spectra more closely, we performed one-dimensional (1D) cuts along the $|Q|$ and energy axes from both the low-temperature and high-temperature data. In the low $|Q|$ range of [0.6, 1.9] {\AA}$^{-1}$ as shown in Fig. \ref{fig2}(e), where the intensity of phonons should be weak, the difference in intensities around 1 meV may be ascribed to temperature-enhanced spin fluctuations followed by the Bose factor. In Fig.~\ref{fig2}(f), we present 1D cuts along the $|Q|$ axis by integrating the energies between 2.5 and 3.5 meV from the map in Fig. \ref{fig2}(b). The cut at 300 K in the lower panel of Fig. \ref{fig2}(f) reveals obvious acoustic phonons, whose intensities become much weaker at 7 K. Interestingly, obvious signals above the background at 7 K can be observed below 3 \AA$^{-1}$ as shown in the upper panel of Fig. \ref{fig2}(f). The intensities observed at low $|Q|$ and low temperature of 7 K should be magnetic in origin.

\begin{figure}[t]
\center{\includegraphics[width=1\linewidth]{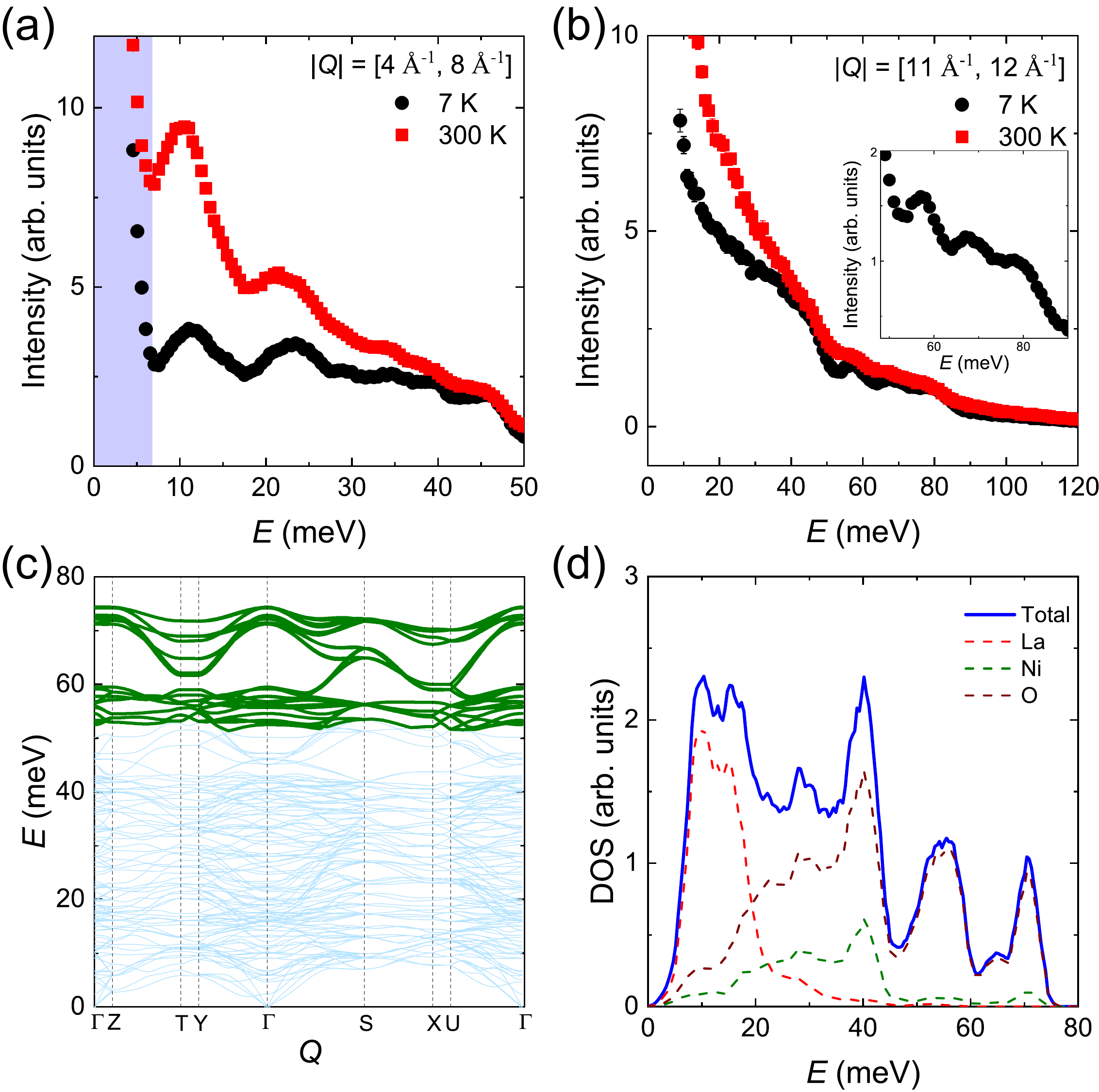}}
\caption{{Phonon modes and vibrational density of states (DOS) of La$_3$Ni$_2$O$_{7-\delta}$.}
(a)-(b) Constant $|Q|$ cuts with the momentum transfer $|Q|$ = [4, 8] {\AA}$^{-1}$ for $E_i$ = 66.3 meV and $|Q|$ = [11, 12] {\AA}$^{-1}$ for $E_i$ = 160 meV. The inset of (b) shows a zoomed-in region of the spectrum at 7 K, revealing peaks around 57, 78, and 80 meV.
(c) DFT calculations of the dispersion of phonons of La$_3$Ni$_2$O$_7$. The dispersive modes in the high-energy region dominated by oxygen are highlighted in olive.
(d) The calculated total vibrational DOS and partial DOS for different elements of La$_3$Ni$_2$O$_7$.}
\label{fig4}
\end{figure}

Since weak magnetic excitations at higher energies may be embedded in the INS spectra, we performed direct data subtractions between 7 and 300 K, as shown in Fig. \ref{fig2} and the SI~\cite{SI}. The subtraction results are shown in Fig.~\ref{fig3}. Phonon intensities at 300 K are expected to be much stronger than at 7 K. However, a flat mode centered at 45 meV appears for the data collected at MERLIN [Figs. \ref{fig3}(a) and \ref{fig3}(b)] and MAPS [Fig.~\ref{fig3}(c)]. The intensities decrease as $|Q|$ increases as shown in the inset of Fig.~\ref{fig3}(a)], indicating the magnetic origin.
We show a 1D cut in the $|Q|$ range of [1.8, 4]~{\AA}$^{-1}$ along the energy axis to see the exact energy of the magnetic excitations in Fig. \ref{fig3}(d). The peak of the magnetic mode centers at 45 meV. The tiny positive intensities around $\sim70-80$~meV are likely to originate from the mismatch of the line width of the optical phonon modes between 7 and 300 K. The $|Q|$ dependence of these intensities does not follow a magnetic form factor. Thus, we only consider the features around 45 meV to be magnetic excitations.

We now attempt to estimate the magnetic exchange couplings by assuming a certain magnetic structure and comparing its magnetic excitation spectrum to the INS data. Although direct evidences on the magnetic structure have never been reported so far, several magnetic structures including the single spin-charge stripe [Fig.~\ref{fig3}(e)], double spin-charge stripe [Fig. S5(c) of SI~\cite{SI}] and double spin stripe [Fig.~\ref{fig3}(g)] have been proposed by recent $\mu$SR, RIXS and NMR studies~\cite{Khasanov2024musR,Chen2024RIXS,Dan2024NMR}. The spin-charge stripe orders require two types of Ni atoms, magnetic and non-magnetic atoms, to form magnetic and non-magnetic stripes. Then the alternated magnetic and non-magnetic stripes give rise to the single spin-charge stripe [Fig.~\ref{fig3}(e)], and the double spin-charge stripe [Fig. S5(c) of SI~\cite{SI}] based on the number of the non-magnetic stripes. The double spin stripe requires all the Ni atoms to be magnetic, and form alternated double stripes with spin-up and spin-down [Fig.~\ref{fig3}(g)]. These three stripe orders were first given in the RIXS study on a single-crystal La$_3$Ni$_2$O$_7$ sample based on the deduced propagation vector ($\bf{k}$ = (0.25, 0.25, 1) in the pseudo-tetragonal notation [$\bf{k}$ = (0.5, 0, 1) in the orthorhombic notation]~\cite{Chen2024RIXS}. The single spin-charge stripe order and the double spin stripe order were demonstrated to reproduce the experimental RIXS results well, while the double spin-charge stripe cannot capture the RIXS spectra well~\cite{Chen2024RIXS}. The double spin stripe order was also proposed by the NMR study on a single-crystal La$_3$Ni$_2$O$_7$ sample since the NMR study did not see the charge ordering that the spin-charge stripe orders require~\cite{Dan2024NMR}. Additionally, the NMR studies also suggested that the tiny ordered magnetic moments ($\textless~0.1~\mu_{\rm B}$) are along the $c$ axis~\cite{Dan2024NMR}. However, the $\mu$SR results from a powder La$_3$Ni$_2$O$_{7-\delta}$ sample excluded all the magnetic structure without charge-only stripes and supported the single spin-charge stripe or double spin-charge stripe orders. Obviously, there are consensus and contradictions in these three studies. The consensuses are the stripe-type magnetic order and the proportion vector $\bf{k}$ = (0.5, 0, 1), while the major disagreement is the specific type of the stripe orders.

With these studies, here we mainly consider these three stripe orders to discuss our INS data. We perform calculations of the spin excitations with the three stripe orders using the linear spin-wave theory (LSWT) method and the SpinW software package. With these LSWT simulations, we find that the single spin-charge stripe order and the double spin stripe orders with appropriate exchange couplings can give rise to spin excitations like that we observed in Figs.~\ref{fig3}(a)-(c). Here we extract the main results of the simulations and present in Figs.~\ref{fig3}(f) and (h). The simulations of the spin excitations of the three magnetic orders with different combinations of the exchange couplings are detailed in the SI~\cite{SI}. We employ a Heisenberg Hamiltonian by considering the effective intralayer exchange couplings $SJ_n$ ($n$ = 1, 2, 3, 4...), and the inter-layer coupling $SJ_{\perp}$ within a bilayer. In Fig.~\ref{fig3}(f), we present the calculated magnetic excitation spectrum using $SJ_2$ = 3.0, $SJ_3$ = 1.7, and $SJ_{\perp}$ = 57.5 meV with the single spin-charge stripe [Fig.~\ref{fig3}(e)]. Similarly, the calculated magnetic excitation spectrum with the double spin stripe order using $SJ_3$ = 3.5, and $SJ_{\perp}$ = 63.9 meV is shown in Fig.~\ref{fig3}(h). The two calculated spectra look very similar and have a common feature that the main spectrum weights of both spectra are around 45 meV, which match the experimental spectra well to some extend. Although the analyses of the experimental and the calculated spectra are cursory, the relatively strong interlayer coupling and weak intralayer couplings should be a key information.

The acoustic phonon branches have been shown in Figs. \ref{fig2}(b) and \ref{fig2}(f). The 1D constant $|Q|$ cuts focusing on the optical phonons are exhibited in Figs.~\ref{fig4}(a) ($E_i$ = 66.3 meV) and~\ref{fig4}(b) ($E_i$ = 160 meV). The most notable features in Fig.~\ref{fig4}(a) are the strong peaks around 11 and 23 meV and weak peaks around 30, 34, and 40 meV, which all have stronger intensities at 300 K than that at 7 K, confirming their phononic origin. Similarly, the INS peaks around 58, 68, and 78 meV that correspond to the flat bands in Figs.~\ref{fig2}(c) can be identified [Fig.~\ref{fig4}(b)]. The phonon calculations were conducted using the DFT method. The dispersion of the phonon modes along the high-symmetry directions is shown in Fig.~\ref{fig4}(c). A large number of phonon branches with different levels of dispersion can be seen. The vibrational density of states is presented in Fig.~\ref{fig4}(d), where the peak features around 10-15, 30, 40, 55, and 70 meV can be recognized. These peaks are roughly consistent with the experimental INS peaks in Figs.~\ref{fig4}(a) and~\ref{fig4}(b). The calculated partial DOS shows that the higher flat bands are dominated by the vibration of oxygen atoms as expected.

\textit{Discussion and Conclusion}.
Our detailed comparisons of the INS spectra at low and high temperatures reveal convincing evidence of magnetic excitations in La$_3$Ni$_2$O$_{7-\delta}$. Although the results are obtained at ambient pressure, the proposed magnetic exchange couplings are important for understanding the mechanism of superconductivity and the pairing gap symmetry in the superconducting state under high pressure~\cite{Yang2023Possible,Zhang2023Trends,Liu2023S+-,Qu2023bilayer}. The results also indicate the existence of spin fluctuations is universal among copper-oxide, iron-based superconductors, and nickelate high $T_c$ superconductors.

We briefly discuss the reasons for the weakness of the spin fluctuations in La$_3$Ni$_2$O$_{7-\delta}$. First of all, the magnetic moments of the Ni atoms here are small~\cite{Khasanov2024musR,Dan2024NMR}, which will reduce the intensity of the spin excitations as well as the magnetic Bragg peaks. Secondly, as shown in Fig.~\ref{fig1}(b), the 3$d_{z^2}$ electrons can form $\sigma$-bonding and anti-bonding bands within a bilayer unit and the 3$d_{x^2-y^2}$ electrons hybridize with the 3$d_{z^2}$ electrons within one Ni-O layer, which give rise to nearly half-filled 3$d_{z^2}$ orbit and quarter-filled 3$d_{x^2-y^2}$ orbit. The quarter-filled character of the 3$d_{x^2-y^2}$ orbit represents a weak intralayer exchange coupling which further reduces the intensity of spin fluctuations. This behavior is different from that in cuprate and iron-based superconductors, where the dominant intralayer exchange couplings lead to the formation of in-plane long-range magnetic order/correlation. Thirdly, polycrystalline samples always give weak signal intensity in INS spectra due to the averaging effect when compared with the results from single-crystal samples with a similar mass. In powder samples of iron-based superconductors, the spin fluctuations are almost invisible~\cite{Qiu2008LaFeAsOF,Christianson2008LaFeAsOF,Mittal2008BaFeAs}, except for the case when there is a strong neutron spin resonance in superconducting samples~\cite{Christianson2008resonance}.

In conclusion, we report the elastic and inelastic neutron scattering results of a polycrystalline La$_3$Ni$_2$O$_{7-\delta}$ sample at ambient pressure. No long-range magnetic order can be identified at 10 K. Our INS results reveal magnetic excitations in La$_3$Ni$_2$O$_{7-\delta}$, indicating that the nickelate superconductors might be in a similar position to the copper-oxide and iron-based superconductors when considering the role of magnetism in the mechanism for high-$T_c$ superconductivity. Estimation the magnitude of the magnetic exchange couplings by assuming the single spin-charge stripe order and the double spin stripe orders results in a strong interlayer coupling and weak intralayer couplings. Our work provides crucial information to the community for determining a realistic mechanism of superconductivity in La$_3$Ni$_2$O$_7$.

\textit{Conflict of interest}: The authors declare that they have no conflict of interest.

\textit{Acknowledgements}. We thank Prof. D. L. Feng for sharing the RIXS data in the `Workshop on High-$T_c$ Nickelate Superconductors'. Work at SYSU was supported by the National Key Research and Development Program of China (Grant Nos. 2023YFA1406500, 2023YFA1406002), the National Natural Science Foundation of China (Grant nos. 12304187, 12174454, U21301001), the Guangdong Basic and Applied Basic Research Funds (grant no. 2021B1515120015), Guangzhou Basic and Applied Basic Research Funds (grant nos. 202201011123, 2024A04J6417), Guangdong Provincial Key Laboratory of Magnetoelectric Physics and Devices (grant no. 2022B1212010008), and the Fundamental Research Funds for the Central Universities, Sun Yat-sen University (Grant No. 23qnpy57). This work is based on experiments (Proposal P15658) performed at Australian Centre for Neutron Scattering (ACNS) at Australian Nuclear Science and Technology Organisation (ANSTO), Australia. The neutron experiment at ISIS Neutron and Muon Source in UK was performed under a user program (Proposal RB2300045 and RB2300079). The neutron diffraction experiment at CSNS in China was performed under the proposal P1822113000042.

\bibliography{La3Ni2O7}
~\\
~\\
\textbf{Supplementary Information: Neutron Scattering Studies on the High-$T_c$ Superconductor La$_3$Ni$_2$O$_{7-\delta}$ at Ambient Pressure}

\section{Neutron Scattering Results}
\begin{figure}[!h]
    \center{\includegraphics[width=.8\linewidth]{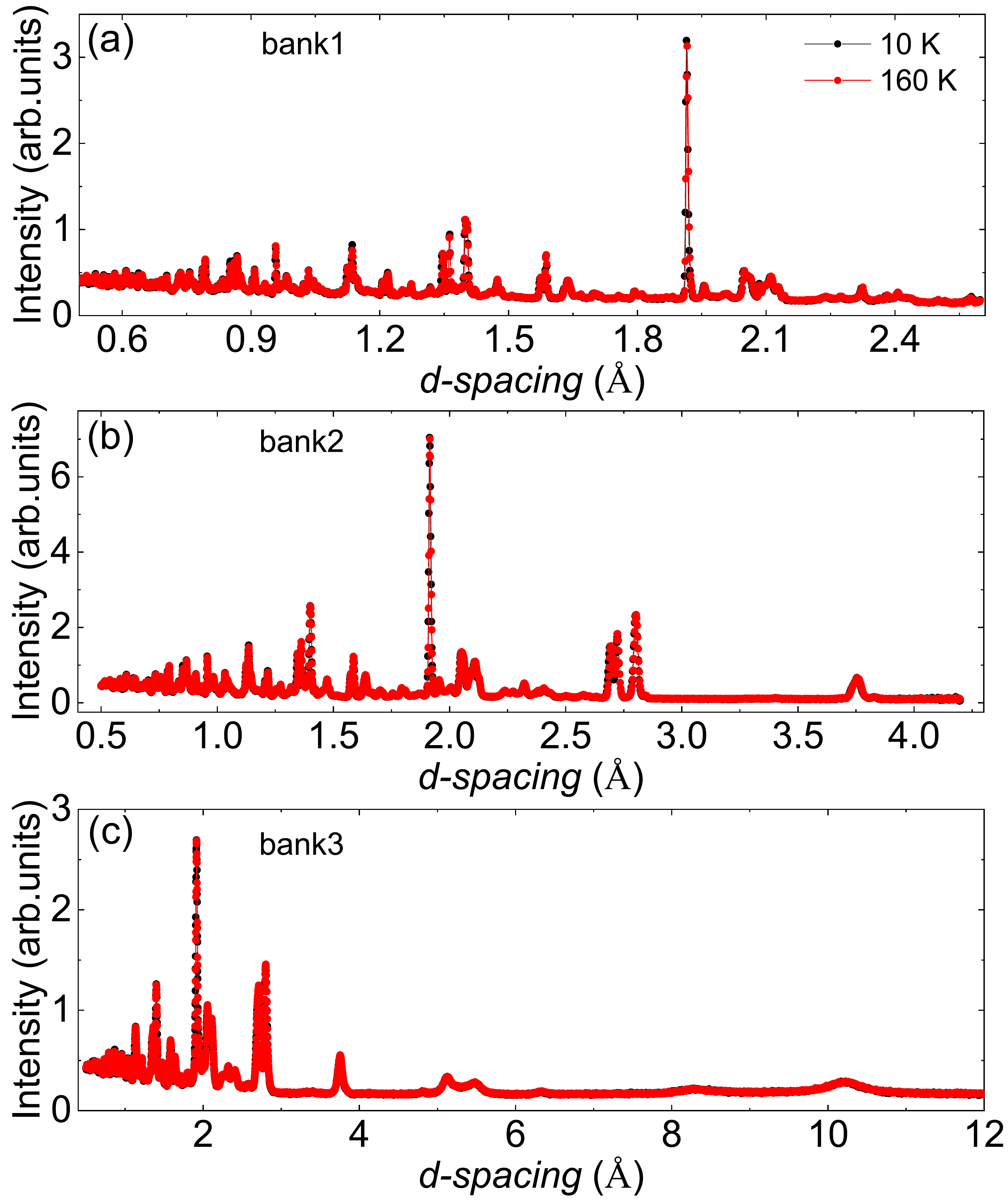}}
    \caption{Neutron powder diffraction results collected on the different detector banks of the GPPD at 10 K and 160 K.}
    \label{FigS_GPPD}
\end{figure}

\begin{figure}[tb]
    \center{\includegraphics[width=.8\linewidth]{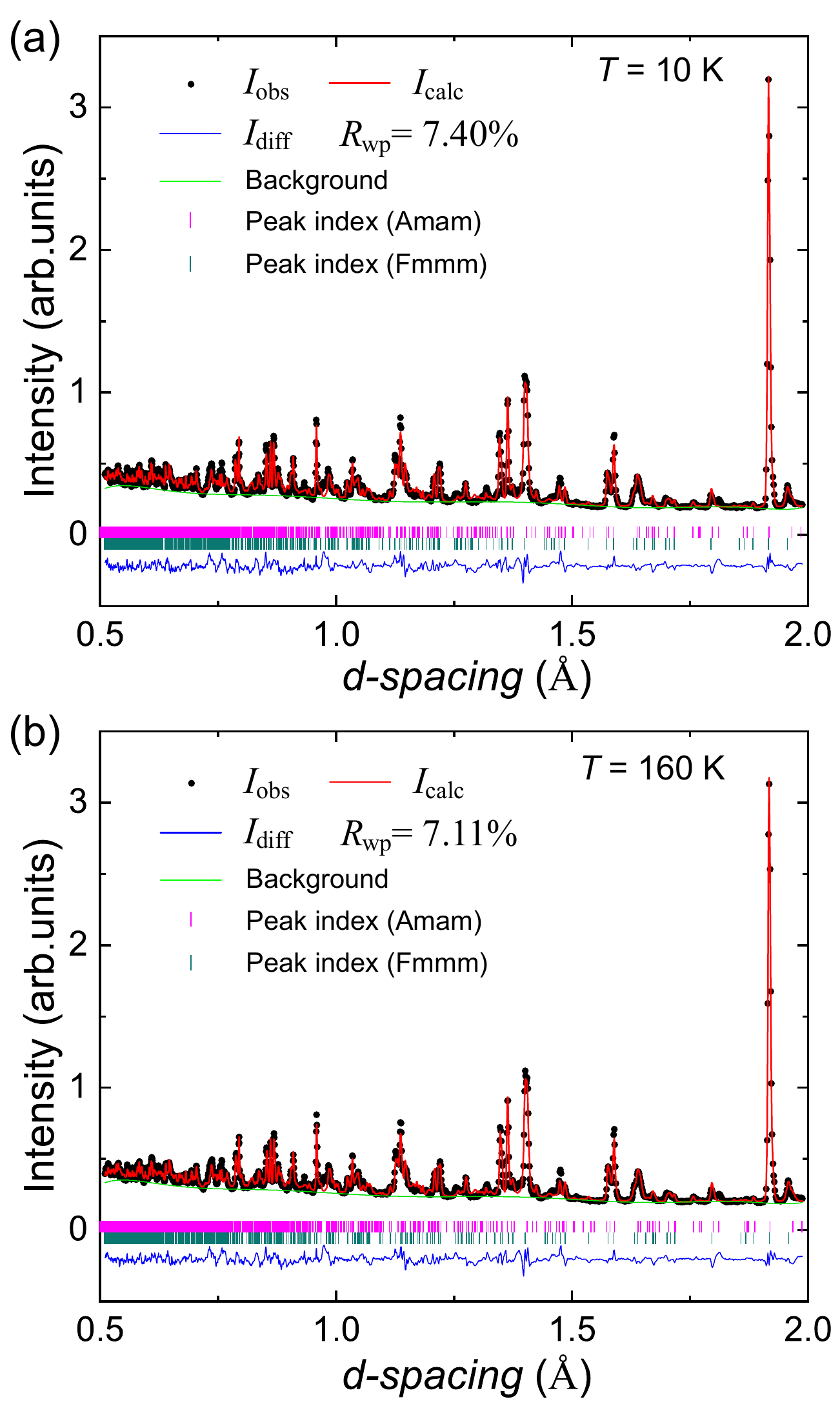}}
    \caption{The Rietveld refinements of the neutron powder diffraction results at 10 K and 160 K.}
    \label{FigS_GPPD_refinement}
\end{figure}

Here we present the neutron scattering raw data that are referred to in the main text. We first compare the neutron powder diffraction results collected on GPPD at 10 K and 160 K in Fig.~\ref{FigS_GPPD}. No new peaks or clear intensity gain on the nuclear peaks from magnetic order can be identified at 10 K. Structural models are refined against the neutron diffraction data with the Rietveld method using the GSAS software package. The refined results are shown in Fig.~\ref{FigS_GPPD_refinement}, and the refined structure parameters at 10 and 160 K are summarized in Tables~\ref{Amam10K},~\ref{Fmmm10K},~\ref{Amam160K},~\ref{Fmmm160K}.

\begin{figure}[t]
    \center{\includegraphics[width=.9\linewidth]{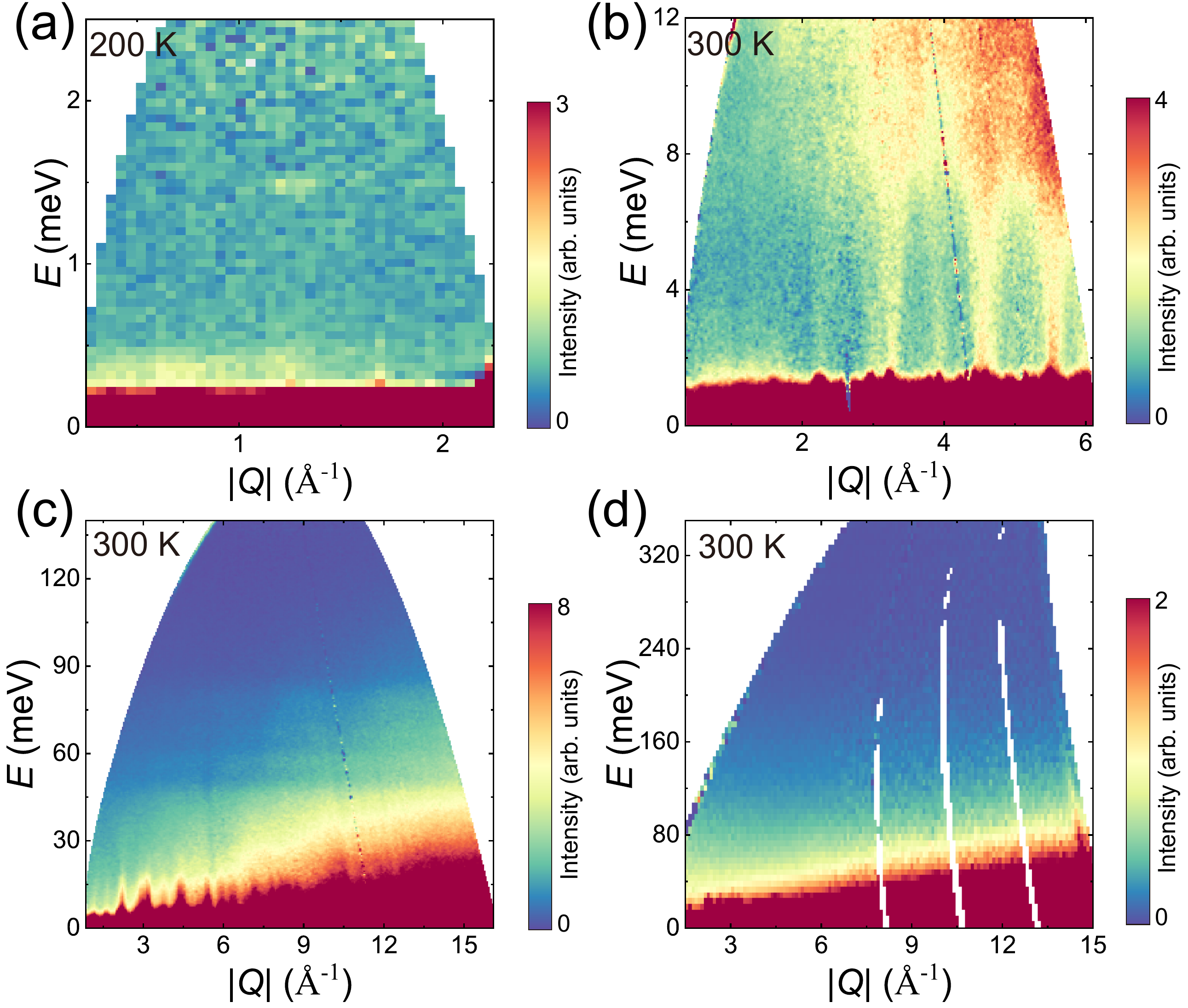}}
    \caption{The corresponding high-temperature INS spectra of La$_3$Ni$_2$O$_{7-\delta}$ with different incident energies to those shown in Fig. 2 of the main text. Panels (a)-(d) are the INS spectra for $E_i$ = 3.7, 22.6, 160, and 350 meV.}
    \label{FigS_INS1}
\end{figure}

\begin{figure}[h]
    \center{\includegraphics[width=.9\linewidth]{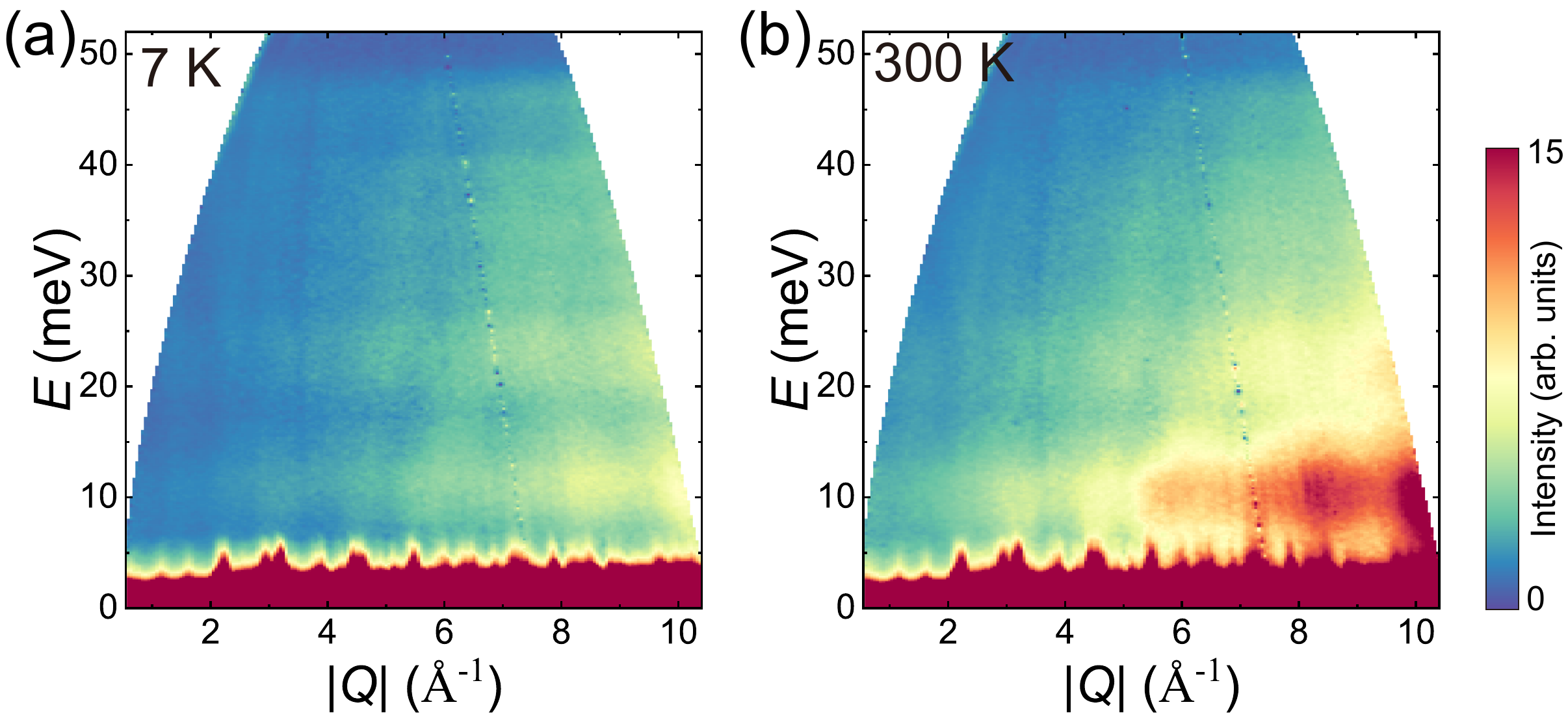}}
    \caption{The raw data used for the 1D cuts in Fig. 4(a) of the main text with $E_i$ = 66.3 meV. }
    \label{FigS_INS2}
\end{figure}

In Fig.~\ref{FigS_INS1}, the corresponding high-temperature INS spectra to those shown in Fig. 2 of the main text are presented. These results were subtracted from the corresponding low-temperature data in Fig. 2 to give the difference shown in Fig. 3 of the main text. In Fig.~\ref{FigS_INS2}, the raw data used for the 1D INS spectra in Fig. 4(a) are presented.

\begin{table*}[!h]
\caption{Refined crystallographic parameters of the $Amam$ phase (No. 63) (56.1\%) at 10 K for our La$_3$Ni$_2$O$_{7-\delta}$ sample.}
\begin{ruledtabular}
\begin{tabular}{c c c c c}
$\;$ Lattice    & $Amam$ (No. 63)     &  $a$ = 5.3956(4)\AA   &  $b$ = 5.4491(4)\AA  &  $c$ = 20.598(3)\AA  \\ \hline \hline
$\;$ atom       & $x$         & $y$          & $z$               & occupancy             \\ \hline
$\;$ La1        & 0.2500     & 0.2453      & 0.3031           & 1                     \\ \hline
$\;$ La2        & 0.2500     & 0.2362      & 0.5000           & 1                    \\ \hline
$\;$ Ni         & 0.2500     & 0.2509      & 0.1059           & 1                    \\ \hline
$\;$ O1         & 0.5000     & 0.0000      & 0.0861           & 1                     \\ \hline
$\;$ O2         & 0.2500     & 0.2126      & 0.1996           & 1                     \\ \hline
$\;$ O3         & 0.0000     & 0.5000      & 0.1055           & 1                     \\ \hline
$\;$ O4         & 0.2500     & 0.3014      & 0.0000           & 0.79(6)                 \\
\end{tabular}
\end{ruledtabular}
\label{Amam10K}
\end{table*}

\begin{table*}[!h]
\caption{Refined crystallographic parameters of the $Fmmm$ phase (No. 69) (41.0\%) at 10 K for our La$_3$Ni$_2$O$_{7-\delta}$ sample.}
\begin{ruledtabular}
\begin{tabular}{c c c c c}
$\;$ Lattice    & $Fmmm$ (No. 69)     &  $a$ = 5.3816(2)\AA   &  $b$ = 5.4538(3)\AA  &  $c$ = 20.470(2)\AA  \\ \hline \hline
$\;$ atom       & $x$         & $y$          & $z$              & occupancy             \\ \hline
$\;$ La1        & 0.0000     & 0.0000      & 0.5000          & 1                     \\ \hline
$\;$ La2        & 0.0000     & 0.0000      & 0.3178          & 1                    \\ \hline
$\;$ Ni         & 0.0000     & 0.0000      & 0.0942          & 1                    \\ \hline
$\;$ O1         & 0.0000     & 0.0000      & 0.0000          & 0.98(8)                \\ \hline
$\;$ O2         & 0.0000     & 0.0000      & 0.2064          & 1                     \\ \hline
$\;$ O3         & -0.2500    & 0.2500      & 0.0934          & 0.98(2)                 \\
\end{tabular}
\end{ruledtabular}
\label{Fmmm10K}
\end{table*}

\begin{table*}[!h]
\caption{Refined crystallographic parameters of the $Amam$ phase (No. 63) (55.8\%) at 160 K for our La$_3$Ni$_2$O$_{7-\delta}$ sample.}
\begin{ruledtabular}
\begin{tabular}{c c c c c}
$\;$ Lattice    & $Amam$ (No. 63)     &  $a$ = 5.4015(4)\AA   &  $b$ = 5.4491(4)\AA  &  $c$ = 20.616(3)\AA  \\ \hline \hline
$\;$ atom       & $x$         & $y$          & $z$               & occupancy             \\ \hline
$\;$ La1        & 0.2500     & 0.2453      & 0.3031           & 1                     \\ \hline
$\;$ La2        & 0.2500     & 0.2362      & 0.5000           & 1                    \\ \hline
$\;$ Ni         & 0.2500     & 0.2509      & 0.1059           & 1                    \\ \hline
$\;$ O1         & 0.5000     & 0.0000      & 0.0861           & 1                     \\ \hline
$\;$ O2         & 0.2500     & 0.2126      & 0.1996           & 1                     \\ \hline
$\;$ O3         & 0.0000     & 0.5000      & 0.1055           & 1                     \\ \hline
$\;$ O4         & 0.2500     & 0.3014      & 0.0000           & 0.81(6)                 \\
\end{tabular}
\end{ruledtabular}
\label{Amam160K}
\end{table*}

\begin{table*}[!h]
\caption{Refined crystallographic parameters of the $Fmmm$ phase (No. 69) (41.3\%) at 160 K for our La$_3$Ni$_2$O$_{7-\delta}$ sample.}
\begin{ruledtabular}
\begin{tabular}{c c c c c}
$\;$ Lattice    & $Fmmm$ (No. 69)     &  $a$ = 5.3880(2)\AA   &  $b$ = 5.4531(3)\AA  &  $c$ = 20.489(2)\AA  \\ \hline \hline
$\;$ atom       & $x$         & $y$          & $z$              & occupancy             \\ \hline
$\;$ La1        & 0.0000     & 0.0000      & 0.5000          & 1                     \\ \hline
$\;$ La2        & 0.0000     & 0.0000      & 0.3178          & 1                    \\ \hline
$\;$ Ni         & 0.0000     & 0.0000      & 0.0942          & 1                    \\ \hline
$\;$ O1         & 0.0000     & 0.0000      & 0.0000          & 0.98(8)                \\ \hline
$\;$ O2         & 0.0000     & 0.0000      & 0.2064          & 1                     \\ \hline
$\;$ O3         & -0.2500    & 0.2500      & 0.0934          & 0.99(2)                 \\
\end{tabular}
\end{ruledtabular}
\label{Fmmm160K}
\end{table*}

\section{Linear Spin-Wave Theory Simulation}
\begin{figure*}[tb]
    \center{\includegraphics[width=.8\textwidth]{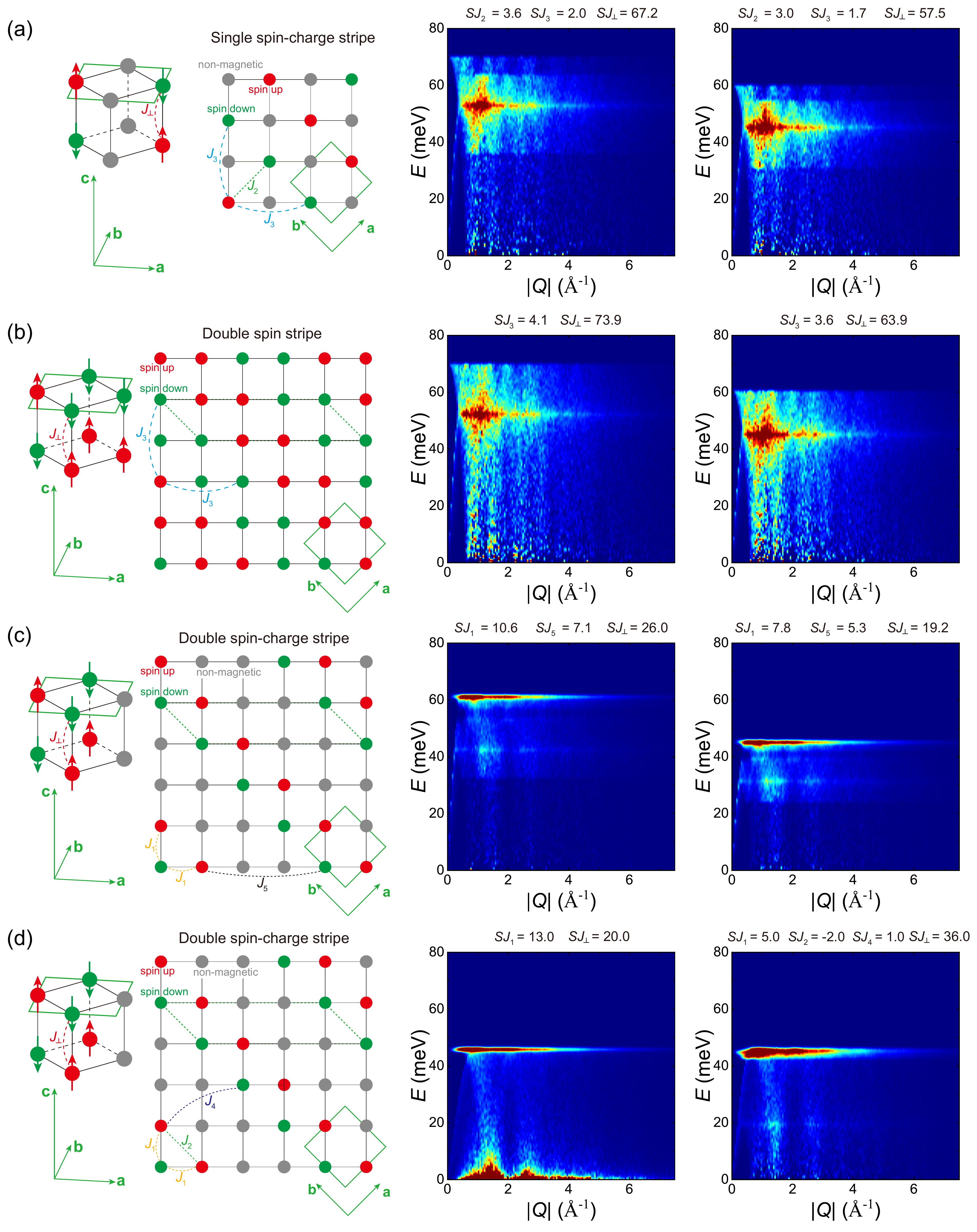}}
    \caption{Different magnetic structures and the corresponding LSWT simulations with different effective exchange parameters for La$_3$Ni$_2$O$_7$. (a) The single spin-charge stripe AFM order with only half of the magnetic Ni ions. (b) The double spin stripe AFM order. (c) and (d) The double spin-charge stripe AFM order with only half of the magnetic Ni ions.}
    \label{FigS2}
\end{figure*}

Since only weak and non-dispersive magnetic fluctuations can be observed in our data, we cannot do quantitative fitting to the data but perform qualitative estimations based on the results. Our strategy to simulate the spin excitations is to assume a magnetic structure and suitable exchange couplings to produce a similar excitation spectrum in the experiment. We note the linear spin-wave theory (LSWT) is suitable for a system with static magnetic order, which is observed in our diffraction experiments. As we mentioned in the main text, we use the proposed magnetic structures from $\mu$SR, RIXS and NMR studies~\cite{Khasanov2024musR,Chen2024RIXS,Dan2024NMR}. We first use the reported exchange coupling parameters in Ref.~\cite{Chen2024RIXS} to calculate the spin excitation spectra. The results are shown in the middle column of Figs.~\ref{FigS2}(a)-(c) for the single spin-charge stripe, double spin stripe, and double spin-charge stripe orders, respectively. The calculations of the single spin-charge stripe and double spin stripe orders look very similar, and the main spectrum weights of both spectra are around 52-53 meV. The calculation of the double spin-charge stripe order produces a rather sharp flat feature around 61-62 meV. Obviously, these calculations from the parameters in Ref.~\cite{Chen2024RIXS} all give main spectrum weights at energies higher than what we observed in INS experiments. Therefore, we used smaller exchange coupling parameters [see that in Figs.~\ref{FigS2}(a)-(c)] than that in Ref.~\cite{Chen2024RIXS} to calculate the spin excitation spectra again. The results are shown in the third column of Figs.~\ref{FigS2}(a)-(c), and the energies of the main spectrum weights around 45 meV now. It should be noted that the flat mode in the calculation of the double spin-charge stripe order is too narrow to be the one we observed. We thus exclude the exchange coupling parameters in Fig.~\ref{FigS2}(c) first. We then try some other cases of the exchange parameters for the double spin-charge stripe order, and two of them are presented in Fig.~\ref{FigS2}(d). Although, the spectrum weight has a certain amount of broadening, it is still too narrow to be the one we observed. We thus exclude the double spin-charge stripe order here. It is worth noting that the double spin-charge stripe order was also excluded by the RIXS data~\cite{Chen2024RIXS}. After reducing the energies of the main spectrum weights to around 45 meV, the calculated spectra in the third column of Figs.~\ref{FigS2}(a)-(b) can both capture the main feature of our INS results. However, the observed spin excitations cannot help to distinguish the single spin-charge stripe and the double spin stripe order.

\end{document}